\documentclass[aps,prd,eqsecnum,preprint,tightenlines,nofootinbib,showpacs]{revtex4-1}
\usepackage{graphicx,amsmath,latexsym}

\def\bea{\begin{eqnarray}}
\def\eea{\end{eqnarray}}
\def\be{\begin{equation}}
\def\ee{\end{equation}}
\newcommand{\ub}[1]{\underline{#1}}
\newcommand{\ob}[1]{\overline{#1}}
\newcommand{\Pminus}{{\cal P}^-}
\newcommand{\Pfree}{{\cal P}_{\rm free}^-}
\newcommand{\Pfreephi}{{\cal P}_{\phi \rm free}^-}
\newcommand{\Pfreechi}{{\cal P}_{\chi \rm free}^-}
\newcommand{\Pint}{{\cal P}_{\rm int}^-}
\newcommand{\pp}{p^{\prime +}}
\newcommand{\ppone}{p_1^{\prime +}}
\newcommand{\pptwo}{p_2^{\prime +}}

\begin{document}

\title{Zero modes in the light-front coupled-cluster method}

\author{Sophia S. Chabysheva}
\author{John R. Hiller}
\affiliation{Department of Physics \\
University of Minnesota-Duluth \\
Duluth, Minnesota 55812}

\date{\today}

\begin{abstract}
The light-front coupled-cluster (LFCC) method is a technique
for solving Hamiltonian eigenvalue problems in light-front-quantized
field theories.  Its primary purpose is to provide a systematic
sequence of solvable approximations to the original
eigenvalue problem without the truncation of Fock space.
Here we discuss the incorporation of zero modes, modes of
zero longitudinal momentum, into the formalism of the method.
Without zero modes, the light-front vacuum is trivial, and the
vacuum expectation value of the field is always zero.  The LFCC method with
zero modes provides for vacuum structure, in the form of a
generalized coherent state of zero modes, as is illustrated here
in two-dimensional model field theories.  
\end{abstract}

%

\maketitle

\section{Introduction}
\label{sec:Introduction}

A very useful approach to the nonperturbative solution of a strongly
interacting quantum field theory is that of Hamiltonian
methods in light-front quantization~\cite{Dirac,DLCQreview}.
In particular, Fock-state expansions of the eigenstates are well-defined,
and there is a separation of external and internal momenta for the constituents,
which leads to boost-invariant wave functions~\cite{DLCQreview}.
Another advantage for most calculations is that
the perturbative vacuum $|0\rangle$ is the physical vacuum;
there is no need to compute the vacuum state before computing
massive eigenstates.  However, this creates a challenging problem
for the calculation of vacuum effects, such as symmetry 
breaking~\cite{Heinzl,Robertson,Hornbostel,Pinsky,Grange,RozowskyThorn,Varyetal,ZeroModes}.
To have contributions to the vacuum requires inclusion of zero modes,
modes of zero longitudinal momentum~\cite{MaskawaYamawaki,Heinzl,Robertson}.

As shown in calculations based on the discrete light-cone quantization
(DLCQ) technique~\cite{PauliBrodsky}, a signal for spontaneous symmetry 
breaking in $\phi^4$ theory can be detected without zero modes by investigation
of a ground-state degeneracy in the massive sector~\cite{RozowskyThorn,Varyetal}.
The discretization can be done with either periodic or antiperiodic boundary
conditions.  In the latter case, zero modes are never present, and in the
former case, they are simply neglected.  As discussed in \cite{ZeroModes},
the neglect of zero modes worsens, but does not prevent, the convergence
of the numerical calculations in the massive sector.  In the case of cubic
scalar theories, where the spectrum is unbounded from below~\cite{Baym},
DLCQ without zero modes is far less successful; detection of the unboundedness
requires careful extrapolation~\cite{Swenson}, whereas inclusion of zero
modes immediately yields the correct result~\cite{ZeroModes}.  

The absence of zero modes also interferes with the calculation of the vacuum 
expectation value and its critical exponent, and, without a vacuum expectation 
value, the Higgs mechanism cannot function; without a zero mode there is no 
constant shift in the field.  A remedy for this is available for DLCQ~\cite{ZeroModes}.
Here we wish to discuss a remedy for the new light-front coupled-cluster (LFCC)
method~\cite{LFCClett,LFCCqed}

The LFCC method was recently developed for the nonperturbative solution of
light-front Hamiltonian problems.  As originally defined~\cite{LFCClett},
it does not explicitly incorporate zero modes and, with respect to spontaneous
symmetry breaking, would be limited to study of degeneracy in the massive
sector of $\phi^4$ theory, just as is DLCQ without zero modes.  The purpose
of this paper is to rectify this deficiency by explicitly including zero modes
in the LFCC method, so that it can be applied to the study of vacuum expectation 
values and the Higgs mechanism.

The mathematical structure of the LFCC method is
closely related to that of the many-body coupled-cluster
method~\cite{CCorigin} used in nuclear physics and
physical chemistry~\cite{CCreviews}.
Some applications of the many-body coupled-cluster method to
field theories in equal-time quantization have been considered~\cite{CC-QFT},
including analysis of symmetry breaking effects in $\phi^4$ theory.
In equal-time quantization, the vacuum structure is explicitly
nontrivial; the vacuum state must be calculated first, with
particle states then built on the vacuum. The situation
is quite different in light-front quantization, where the
vacuum appears trivial, until zero modes are included.

We include zero modes in the LFCC method by a limiting
procedure, with modes of infinitesimal momentum introduced at the 
start of a calculation and the limit of zero momentum taken at or near
the end.  The vacuum eigenstate then becomes a generalized
coherent state of zero modes~\cite{HariVary,coherentstates}.
The technique is developed in a series of two-dimensional
examples; we discuss $\phi^3$ theory~\cite{VaryHari-phi3}, 
$\phi^4$ theory~\cite{RozowskyThorn,Varyetal,VaryHari-phi4}, and
the Wick--Cutkosky model~\cite{WC}.  In each case, we compute the
energy density of the vacuum and demonstrate the existence of
broken-symmetry solutions at minima in the energy density.
Where possible, we compare these results with a variational
coherent-state analysis.

An overview of the LFCC method is provided in Sec.~\ref{sec:LFCC},
as a precursor to the consideration of zero modes.  The formalism
for zero modes is developed in Sec.~\ref{sec:phi3}, in the context
of $\phi^3$ theory, and then extended to $\phi^4$ theory in Sec.~\ref{sec:phi4}
and the Wick--Cutkosky model in Sec.~\ref{sec:WC}.  Some details
of the calculations in the Wick--Cutkosky model are left to an Appendix.
A brief summary is given in Sec.~\ref{sec:summary}.

\section{Light-front coupled-cluster method}
\label{sec:LFCC}

The light-front Hamiltonian eigenvalue problem is formulated in
Fock space as the fundamental equation
\be \label{eq:eigenvalueproblem}
\Pminus|\psi(\ub{P})\rangle=\frac{M^2+P_\perp^2}{P^+}|\psi(\ub{P})\rangle,
\ee
where $\Pminus$ is the light-front energy operator, conjugate
to the light-front time $x^+\equiv t+z$, and 
$\ub{P}=(P^+\equiv E+P^z,\vec P_\perp=(P^x,P^y))$
is the light-front momentum, with $P^+$ conjugate to $x^-\equiv t-z$.
The eigenstate $|\psi\rangle$ has mass $M$ and momentum $\ub{P}$,
and is expanded in a Fock basis of eigenstates of $\ub{P}$ and 
of particle number.  The coefficients of
the Fock states are the wave functions that describe the
eigenstate.  The eigenvalue problem (\ref{eq:eigenvalueproblem})
is equivalent to an
infinite coupled system of integral equations for these
wave functions.

To have a finite calculation for an eigenstate, the usual step is a truncation
of Fock space, to have a finite number of wave functions and
a finite set of equations.  This, however, leads to many
difficulties, particularly uncanceled divergences~\cite{SecDep}.
The LFCC method~\cite{LFCClett}
is designed to avoid these difficulties by not truncating Fock space
but instead restricting the relationships between wave
functions in such a way as to produce a finite set of
(nonlinear) equations.

The LFCC method constructs the eigenstate $|\psi\rangle$ from
a valence state $|\phi\rangle$, with the smallest number of
constituents, and the exponentiation of an operator $T$ that
increases the particle number, to generate higher Fock states.
The general form is
\be
|\psi(\ub{P})\rangle=\sqrt{Z}e^T|\phi(\ub{P})\rangle,
\ee
with $\sqrt{Z}$ a normalization factor.  The eigenvalue
problem is then converted to a valence eigenvalue problem
\be
P_v \ob{\Pminus}|\phi(\ub{P})\rangle=P^-|\phi(\ub{P})\rangle
    =\frac{M^2+P_\perp^2}{P^+}|\phi(\ub{P})\rangle,
\ee
with $\ob{\Pminus}\equiv e^{-T}\Pminus e^T$ an effective Hamiltonian
and $P_v$ a projection onto the valence sector, and to an auxiliary
equation for $T$, as a projection onto all higher Fock states.
\be \label{eq:aux}
(1-P_v)\ob{\Pminus}|\phi(\ub{P})\rangle=0.
\ee
The auxiliary equation is actually an infinite set of equations for
the infinite set of terms in $T$, and as such we still have an exact representation of
the original eigenvalue problem.  The approximations that lead
to a finite set of equations, without truncating Fock space, are
the truncation of $T$ to a finite number of terms and a matching
truncation of the projection $1-P_v$, to generate the finite number
of equations\footnote{To not truncate $1-P_v$ would lead to an 
overdetermined system of equations for the terms in $T$.}
needed to solve for the terms in $T$.

The structure of the $T$ operator is such that one can include
terms with zero modes.  Such terms allow for
zero-mode contributions to the eigenstates, in particular the
vacuum.  We show this by example, beginning with
$\phi^3$ theory in the next section.

\section{$\phi^3$ theory}
\label{sec:phi3}

\begin{figure}[ht]
\vspace{0.2in}
\begin{center}
\begin{tabular}{c}
\includegraphics[width=8cm]{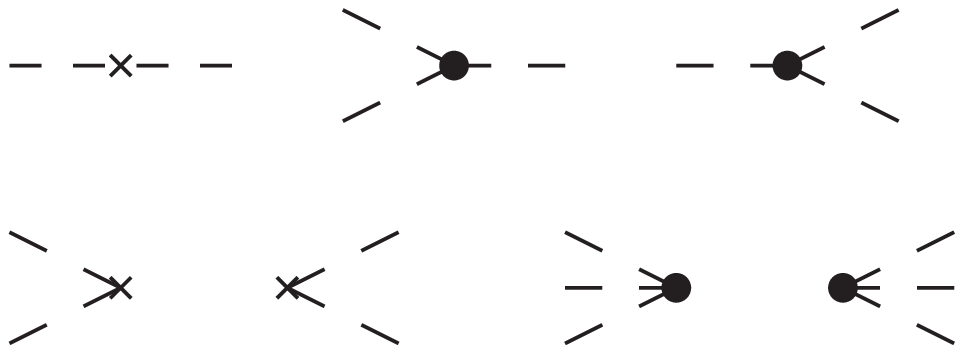} \\
(a) \\
\mbox{} \\
\includegraphics[width=2.5cm]{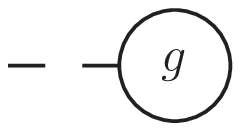} \\
(b)
\end{tabular}
\end{center}
\caption{\label{fig:phi3PandT} 
Diagrammatic representation of (a) the light-front Hamiltonian $\Pminus$
and (b) the approximate $T$ operator for $\phi^3$ theory.  The cross represents
the kinetic energy contribution.  External lines on the right represent
annihilation operators; those on the left, creation operators.
}
\end{figure}

The Lagrangian of $\phi^3$ theory is
\be
{\cal L}=\frac12(\partial_\mu\phi)^2-\frac12\mu^2\phi^2-\frac{\lambda}{3!}\phi^3.
\ee
From this, the two-dimensional light-front Hamiltonian density is
\be
{\cal H}=\frac12 \partial_-\phi \partial_+\phi-{\cal L}
  =\frac12 \mu^2 \phi^2+\frac{\lambda}{3!}\phi^3.
\ee
The mode expansion for the field at zero light-front time is
\be \label{eq:mode}
\phi=\int \frac{dp^+}{\sqrt{4\pi p^+}}
   \left\{ a(p^+)e^{-ip^+x^-/2} + a^\dagger(p^+)e^{ip^+x^-/2}\right\},
\ee
with the modes quantized such that 
\be
[a(p^+),a^\dagger(\pp)]=\delta(p^+-\pp).
\ee
The normal-ordered light-front Hamiltonian $\Pminus=\Pfree+\Pint$ is
then specified by
\bea \label{eq:Pfree}
\Pfree&=&\int dp^+ \frac{\mu^2}{p^+} a^\dagger(p^+)a(p^+) \\
 && +\frac{\mu^2}{2}\int \frac{dp_1^+ dp_2^+}{\sqrt{p_1^+ p_2^+}}
    \delta(p_1^++p_2^+)\left[a^\dagger(p_1^+)a^\dagger(p_2^+)+a(p_1^+)a(p_2^+)\right]
    \nonumber
\eea
and
\bea
\Pint&=&\frac{\lambda}{2}\int \frac{dp^+d\pp}
                              {\sqrt{4\pi p^+ \pp(p^+-\pp)}}
   \left[a^\dagger(p^+)a(\pp)a(p^+-\pp) \right.\\
   && \rule{2in}{0mm} \left.
       +a^\dagger(\pp)a^\dagger(p^+-\pp)a(p^+)\right]  \nonumber \\
 && +\frac{\lambda}{6}\int\frac{dp_1^+ dp_2^+ dp_3^+}{\sqrt{4\pi p_1^+ p_2^+ p_3^+}}
     \delta(p_1^+ + p_2^+ + p_3^+) \nonumber \\
  && \rule{1in}{0mm} \times   \left[ a^\dagger(p_1^+) a^\dagger(p_2^+) a^\dagger(p_3^+)
            +a(p_1^+) a(p_2^+) a(p_3^+)\right]. \nonumber
\eea
The terms with only creation or annihilation operators are usually dropped
in light-front quantization, because each $p^+$ is positive and the delta
functions only have support at $p^+=0$.  Here, however, these terms are
kept as zero-mode contributions.  For a graphical representation of
$\Pminus$, see Fig.~\ref{fig:phi3PandT}.
%
%
The effective Hamiltonian $\ob{\Pminus}$ of the LFCC method is computed
from the Baker--Hausdorff expansion
\be \label{eq:BH}
\ob{\Pminus}=e^{-T}\Pminus e^T=\Pminus+[\Pminus,T]+\frac{1}{2!}[[\Pminus,T],T]+\cdots,
\ee
given an approximation for $T$.

To consider the lowest order zero-mode contribution, we truncate the $T$ operator
to the creation of one zero mode
\be  \label{eq:T0}
T=\int_0^\infty dp^+ \sqrt{4\pi p^+}g(p^+)a^\dagger(p^+),
\ee
with $g(p^+)$ having support only at $p^+=0$ in a limit appropriate for the 
form chosen for $g$.  For example, $g$ could be an exponential $\frac{1}{\epsilon}e^{-p^+/\epsilon}$
or a step function $\frac{1}{\epsilon}\theta(\epsilon-p^+)$,
with the appropriate limit being $\epsilon\rightarrow0$.
This limit, taken at the end of the calculation, restores momentum
conservation.  The valence state is the bare vacuum, and the
projection $1-P_v$ is truncated to include only states with one zero mode.
The corresponding transformation of the field is
\be \label{eq:shift}
e^{-T}\phi e^T=\phi+[\phi,T]+\cdots=\phi +\int dp^+ g(p^+) e^{-ip^+x^-/2},
\ee
which provides for a constant shift in the limit that $g(p^+)\propto\delta(p^+)$.
A graphical representation of $T$ is given in Fig.~\ref{fig:phi3PandT}.

A calculation of the terms in the Baker--Hausdorff expansion (\ref{eq:BH})
then determines the effective Hamiltonian.  Only a finite number of
terms will contribute to the eigenvalue problem, because we need only
terms that change the particle number by no more than one; any more
than this would go beyond the truncation of the projection $(1-P_v)$.
We will consider only the vacuum as the valence
state, and, therefore, terms in $\ob{\Pminus}$ with any
annihilation operators will also be neglected; however, such terms
do need to be kept in intermediate calculations of
commutators, because higher-order commutators can reduce the
total number of annihilation operators.

We compute the following commutators for $T$ with $\Pfree$:
\bea \label{eq:PfreeT}
[\Pfree,T]&=&\sqrt{4\pi}\mu^2\int dp^+ \frac{g(p^+)}{\sqrt{p^+}}a^\dagger(p^+) \\
&& +\sqrt{4\pi}\mu^2\int\frac{dp_1^+dp_2^+}{\sqrt{p_1^+}}
                \delta(p_1^++p_2^+)g(p_2^+)a(p_1^+), \nonumber
\eea
\be \label{eq:PfreeT2}
{[}[\Pfree,T],T]=4\pi\mu^2\int dp_1^+ dp_2^+ \delta(p_1^++p_2^+)g(p_1^+)g(p_2^+),
\ee
\be
{[}[[\Pfree,T],T],T]=0,
\ee
and with $\Pint$:
\bea
{[}\Pint,T]&=&\lambda\int\frac{dp^+d\pp}{\sqrt{p^+(p^+-\pp)}}
           g(\pp)a^\dagger(p^+)a(p^+-\pp) \\
      && +\frac{\lambda}{2}\int\frac{dp^+d\pp}{\sqrt{p^+(p^+-\pp)}}
          g(p^+)a^\dagger(\pp)a^\dagger(p^+-\pp) \nonumber \\
      && +\frac{\lambda}{2}\int \frac{dp_1^+ dp_2^+ dp_3^+}{\sqrt{p_1^+ p_2^+}}
          \delta(p_1^+ + p_2^+ + p_3^+) g(p_3^+) a(p_1^+) a(p_2^+), \nonumber 
\eea
\bea
{[}[\Pint,T],T]&=&\sqrt{4\pi}\lambda\int \frac{dp^+ d\pp}{\sqrt{p^+}}
       g(\pp) g(p^+-\pp) a^\dagger(p^+) \\
       && + \sqrt{4\pi}\lambda 
         \int\frac{dp_1^+ dp_2^+ dp_3^+}{\sqrt{p_1^+}}
     \delta(p_1^+ + p_2^+ + p_3^+) g(p_2^+)g(p_3^+) a(p_1^+), \nonumber
\eea
\be
{[}[[\Pint,T],T],T]=4\pi\lambda \int dp_1^+ dp_2^+ dp_3^+
     \delta(p_1^+ + p_2^+ + p_3^+) g(p_1^+)g(p_2^+)g(p_3^+),
\ee
\be
{[}[[[\Pint,T],T],T],T]=0.
\ee
Each commutator with $\Pminus$ contracts one zero-mode creation operator
with one annihilation operator in $\Pminus$.

\begin{figure}[ht]
\vspace{0.2in}
\centerline{\includegraphics[width=14cm]{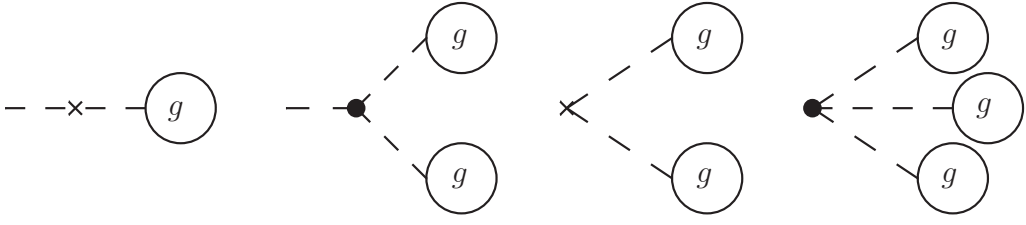}}
\caption{\label{fig:phi3EffP} 
Same as Fig.~\protect\ref{fig:phi3PandT}, but for the zero-mode
terms in the effective Hamiltonian $\ob{\Pminus}$.  Internal lines
represent contractions.
}
\end{figure}

From these commutators we build the expression for $\ob{\Pminus}$,
keeping only those terms that do not annihilate the vacuum and
create at most one zero mode.  This gives
\bea  \label{eq:effP-}
\ob{\Pminus}&=& \sqrt{4\pi}\mu^2\int dp^+ \frac{g(p^+)}{\sqrt{p^+}}a^\dagger(p^+) \\
  && +\frac{1}{2!}4\pi\mu^2\int dp_1^+ dp_2^+ \delta(p_1^++p_2^+)g(p_1^+)g(p_2^+) 
  \nonumber \\
  && +\frac{1}{2!}\sqrt{4\pi}\lambda\int \frac{dp^+ d\pp}{\sqrt{p^+}}
       g(\pp) g(p^+-\pp) a^\dagger(p^+) \nonumber \\
  && +\frac{1}{3!}4\pi\lambda \int dp_1^+ dp_2^+ dp_3^+
     \delta(p_1^+ + p_2^+ + p_3^+) g(p_1^+)g(p_2^+)g(p_3^+). \nonumber
\eea
A graphical representation is given in Fig.~\ref{fig:phi3EffP}
%
%

For the vacuum valence state $|0\rangle$, the
eigenvalue problem in the valence sector
$P_v\ob{\Pminus}|0\rangle=P^-|0\rangle$ is
\bea \label{eq:phi3valence}
\lefteqn{\left[ \frac12 4\pi\mu^2\int dp_1^+ dp_2^+ \delta(p_1^++p_2^+)g(p_1^+)g(p_2^+)
                 \right.}&& \\
&&  \left.+ \frac16 4\pi\lambda \int dp_1^+ dp_2^+ dp_3^+
     \delta(p_1^+ + p_2^+ + p_3^+) g(p_1^+)g(p_2^+)g(p_3^+) \right]|0\rangle
      =P^-|0\rangle.  \nonumber
\eea
If one assumes that $g$ is a function with support only at $p^+=0$, that is 
$g(p^+)=\alpha\delta(p^+)$, the eigenvalue $P^-$ is 
\be
P^-=\frac12 4\pi \mu^2 \alpha^2 \delta(0) +\frac16 4 \pi \lambda \alpha^3 \delta(0).
\ee
The $\delta(0)$ factors are no surprise, because we expect
$P^-$ to be infinite, proportional to the volume
\be
\lim_{p^+\rightarrow0} \int dx^- e^{ip^+x^-/2}=\int dx^- =4\pi\delta(0).
\ee
Therefore, we write $P^-$ in terms of an energy density ${\cal E}^-$ as
\be
P^-={\cal E}^-\int dx^-,
\ee
and find
\be
{\cal E}^-=\frac12 \mu^2 \alpha^2 +\frac16 \lambda \alpha^3.
\ee
Clearly, the spectrum is unbounded from below~\cite{Baym} as $\alpha$ goes
to negative infinity.

The form of the function $g$ was already assumed to be a delta function.
Now let us see that this is exactly what the auxiliary equation
(\ref{eq:aux}) gives.  Truncated to states with only one zero mode,
the auxiliary equation becomes
\be \label{eq:phi3aux}
\sqrt{4\pi}\mu^2 \frac{g(p^+)}{\sqrt{p^+}}
+\frac12\sqrt{4\pi}\lambda\int_0^{p^+} \frac{d\pp}{\sqrt{p^+}}
       g(\pp) g(p^+-\pp) =0.
\ee
This equation can be solved by taking the Laplace transform after
multiplication by $\sqrt{p^+}$.  With the definition
\be
G(s)\equiv \int_0^\infty e^{-sp^+}g(p^+)dp^+,
\ee
Eq.~(\ref{eq:phi3aux}) becomes
\be
\mu^2G(s)+\frac12\lambda G(s)^2=0,
\ee
where the Laplace transform of the convolution in the second term
is just the product of the transforms.
The possible solutions are $G(s)$=0 and $-2\mu^2/\lambda$.  Because
the inverse transform of a constant is a delta function, we
obtain the expected $g(p^+)=\alpha\delta(p^+)$ with
$\alpha=0$ or $\alpha=-2\mu^2/\lambda$.
These are the local extrema of ${\cal E}^-$; the LFCC auxiliary
equation does miss the global extrema at $\pm\infty$.

This analysis leads to a natural choice for a limiting
form to use in the construction of zero-mode $T$ operators.
We define
\be \label{eq:Delta}
\Delta(p^+)=\frac{1}{\epsilon}e^{-p^+/\epsilon}
\ee
so that $\lim_{\epsilon\rightarrow0}\Delta(p^+)=\delta(p^+)$,
for integrals from zero to infinity, and the Laplace transform is 
\be
\int_0^\infty dp^+ e^{-sp^+}\Delta(p^+)=\frac{1}{\epsilon}\frac{1}{s+1/\epsilon}\rightarrow1
\ee
For $\phi^3$ we would then define the truncated $T$ operator as
\be  \label{eq:T-Delta}
T=\alpha\int_0^\infty dp^+ \sqrt{4\pi p^+}\Delta(p^+)a^\dagger(p^+).
\ee

For comparison with the LFCC result, we consider a variational coherent-state
analysis~\cite{coherentstates}
of the light-front vacuum energy density\footnote{The light-front momentum
of the vacuum is, of course, zero.  Thus, the light-front energy density 
is proportional to the ordinary energy density.} 
$\langle\alpha|\!:\!{\cal H}\!:\!|\alpha\rangle$, with respect to a
vacuum state $|\alpha\rangle$.
This provides a direct correspondence with the LFCC result
when, as above, the $T$ operator is truncated to one zero mode, because
$|\alpha\rangle\equiv\sqrt{Z_\alpha}e^T|0\rangle$ is then a coherent state. 
With $T$ represented as in (\ref{eq:T-Delta}), the following commutators
can be computed:
\bea
{[}T^\dagger,T]&=&4\pi\alpha^2\int dp^+ p^+ \Delta^2(p^+)\rightarrow \pi\alpha^2,
\label{eq:TdaggerT} \\
{[}\phi,T]&=&\alpha\int dp^+ \Delta(p^+) e^{-ip^+x^-/2}\rightarrow \alpha, \\
{[}\phi,T^\dagger]&=&\alpha\int dp^+ \Delta(p^+) e^{+ip^+x^-/2}\rightarrow \alpha.
\eea
We then have, for real $\alpha$, $\sqrt{Z_\alpha}=e^{-\pi\alpha^2/2}$, 
$\phi|\alpha\rangle =\alpha|\alpha\rangle$,
$\langle\alpha|\phi=\langle\alpha|\alpha$, and
\be
\langle\alpha|\!:\!{\cal H}\!:\!|\alpha\rangle=\frac12\mu^2\alpha^2+\frac16\lambda\alpha^3={\cal E}^-.
\ee
The local extrema are at $\alpha=0$ and $\alpha=-2\mu^2/\lambda$,
as in the LFCC analysis, and the global extrema at $\pm\infty$.
The vacuum expectation value for the field is just 
$\langle\alpha|\phi|\alpha\rangle=\alpha$.

A different choice for the function $\Delta(p^+)$ would change the result in (\ref{eq:TdaggerT})
for the commutator $[T^\dagger,T]$.  For example, the step function 
$\frac{1}{\epsilon}\theta(\epsilon-p^+)$ would yield
\be
4\pi\alpha^2\int_0^\epsilon \frac{1}{\epsilon^2} p^+ dp^+=2\pi\alpha^2,
\ee
which differs by a factor of two.  However, this changes the relative
normalization $Z_\alpha$ but not the expectation value; 
the commutators $[\phi,T]$ and $[\phi,T^\dagger]$ are unaffected, 
because all forms of $\Delta(p^+)$ must be consistent with
$\delta(p^+)$ as the limit.

\section{$\phi^4$ theory}
\label{sec:phi4}

\begin{figure}[ht]
\vspace{0.2in}
\begin{center}
\begin{tabular}{c}
\includegraphics[width=10cm]{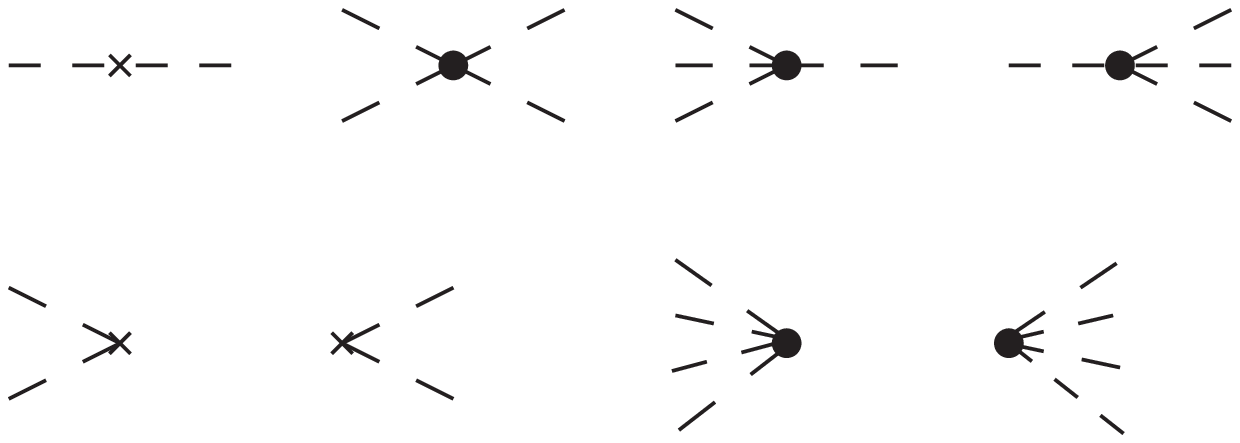} \\
(a) \\
\mbox{} \\
\includegraphics[width=2.5cm]{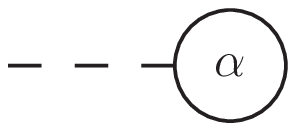} \\
(b)
\end{tabular}
\end{center}
\caption{\label{fig:phi4PandT} 
Same as Fig.~\protect\ref{fig:phi3PandT}, but for $\phi^4$ theory.}
\end{figure}

The Lagrangian and light-front Hamiltonian density for $\phi^4$ theory are
\be
{\cal L}=\frac12(\partial_\mu\phi)^2-\frac12\mu^2\phi^2-\frac{\lambda}{4!}\phi^4
\ee
and
\be
{\cal H}=\frac12 \mu^2 \phi^2+\frac{\lambda}{4!}\phi^4.
\ee
The mode expansion for the field $\phi$ is the same as
(\ref{eq:mode}) for $\phi^3$ theory.
We again split the light-front Hamiltonian $\Pminus$ into two parts,
$\Pfree$, which is given in Eq.~(\ref{eq:Pfree}), and
\bea
\Pint&=&\frac{\lambda}{6}\int \frac{dp_1^+dp_2^+dp_3^+}
                              {4\pi \sqrt{p_1^+p_2^+p_3^+(p_1^++p_2^++p_3^+)}} \\
  &&\rule{1in}{0mm} 
    \times \left[a^\dagger(p_1^++p_2^++p_3^+)a(p_1^+)a(p_2^+)a(p_3^+)\right. \nonumber \\
  && \rule{1.25in}{0mm} \left. 
    +a^\dagger(p_1^+)a^\dagger(p_2^+)a^\dagger(p_3^+)a(p_1^++p_2^++p_3^+)\right]  \nonumber \\
 && +\frac{\lambda}{4}\int\frac{dp_1^+ dp_2^+}{4\pi\sqrt{p_1^+p_2^+}}
       \int\frac{dp_1^{\prime +}dp_2^{\prime +}}{\sqrt{p_1^{\prime +} p_2^{\prime +}}} 
       \delta(p_1^+ + p_2^+-p_1^{\prime +}-p_2^{\prime +}) \nonumber \\
 && \rule{2in}{0mm} \times a^\dagger(p_1^+) a^\dagger(p_2^+) a(p_1^{\prime +}) a(p_2^{\prime +}) 
   \nonumber \\
 && +\frac{\lambda}{24}\int\frac{dp_1^+ dp_2^+ dp_3^+ dp_4^+}{4\pi \sqrt{p_1^+ p_2^+ p_3^+p_4^+}}
     \delta(p_1^+ + p_2^+ + p_3^++p_4^+) \nonumber \\
  && \rule{0.5in}{0mm} \times   
  \left[ a^\dagger(p_1^+) a^\dagger(p_2^+) a^\dagger(p_3^+)a^\dagger(p_4^+)
            +a(p_1^+) a(p_2^+) a(p_3^+) a(p_4^+)\right]. \nonumber
\eea
A graphical representation is shown in Fig.~\ref{fig:phi4PandT}.
%
%

We focus on the zero-mode contributions to a vacuum valence state
and consider the $T$ operator (\ref{eq:T-Delta}) for a single
zero mode.  The variational coherent-state approach gives
\be \label{eq:phi4var}
\langle\alpha|\!:\!{\cal H}\!:\!|\alpha\rangle=\frac12\mu^2\alpha^2+\frac{1}{24}\lambda\alpha^4,
\ee
with a local maximum at $\alpha=0$ and local minima at
$\alpha^2=-6\mu^2/\lambda$.  Of course, the latter is 
realizable only for $\mu^2<0$.

\begin{figure}[ht]
\vspace{0.2in}
\centerline{\includegraphics[width=12cm]{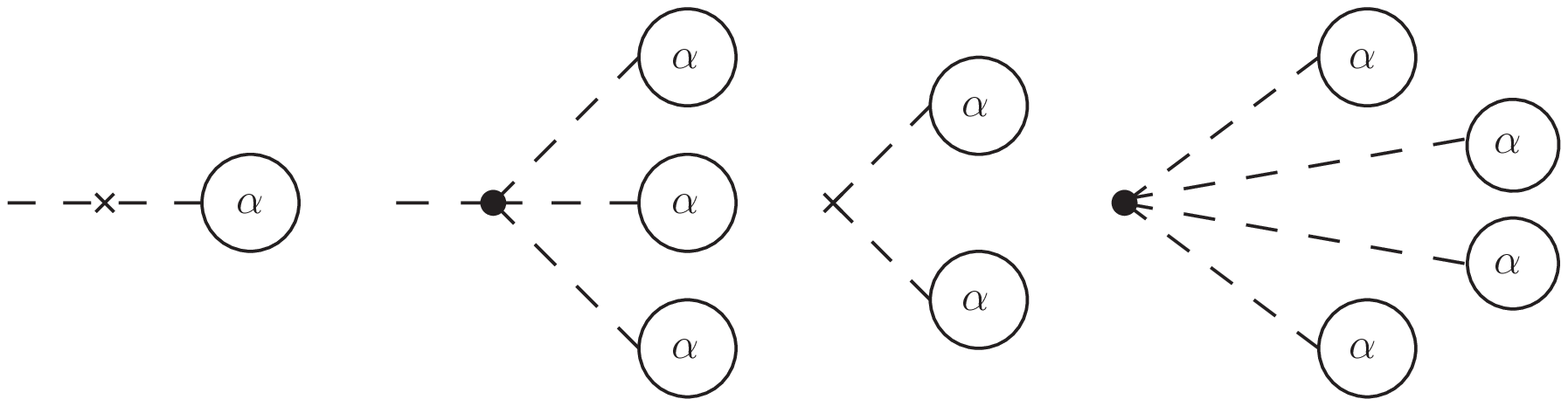}}
\caption{\label{fig:phi4EffP} 
Same as Fig.~\protect\ref{fig:phi3EffP}, but for $\phi^4$ theory.}
\end{figure}

For the LFCC analysis, the relevant commutators with $\Pfree$ are 
the same as those for $\phi^3$ theory, with $g(p^+)=\alpha\Delta(p^+)$,
given in Eqs.~(\ref{eq:PfreeT}) and (\ref{eq:PfreeT2}).  Those for $\Pint$ are
\bea
\label{eq:PintT}
{[}\Pint,T]&=& \frac{\lambda\alpha}{2}\int 
   \frac{dp_1^+ dp_2^+ dp^+}{\sqrt{4\pi p_1^+ p_2^+ p^+}}
     \delta(p^+-p_1^+ -p_2^+) \\
 && \rule{0.5in}{0mm} \times
     \left[a^\dagger(p^+)a(p_1^+) a(p_2^+)
                   +a^\dagger(p_1^+) a^\dagger(p_2^+) a(p^+)\right] \nonumber\\
 && +\frac{\lambda\alpha}{6}\int 
     \frac{dp_1^+ dp_2^+ dp_3^+}{\sqrt{4\pi p_1^+ p_2^+ p_3^+}}
     \delta(p_1^+ +p_2^+ +p_3^+) \nonumber \\
 && \rule{0.5in}{0mm} \times
     \left[ a(p_1^+) a(p_2^+) a(p_3^+) 
         + a^\dagger(p_1^+) a^\dagger(p_2^+) a^\dagger(p_3^+\right], \nonumber
\eea
\bea
{[}[\Pint,T],T]&=&\lambda\alpha^2 \int \frac{dp^+}{p^+} a^\dagger(p^+)a(p^+) \\
  && + \frac{\lambda\alpha^2}{2}\int \frac{dp_1^+ dp_2^+}{\sqrt{p_1^+ p_2^+}}
    \delta(p_1^+ + p_2^+) \left[ a(p_1^+)a(p_2^+)+a^\dagger(p_1^+)a^\dagger(p_2^+)\right],
                 \nonumber
\eea
\be \label{eq:PintT3}
{[}[[\Pint,T],T],T]= \sqrt{4\pi}\lambda\alpha^3\int\frac{dp^+}{\sqrt{p^+}}\Delta(p^+)
                      \left[a(p^+) + a^\dagger(p^+)\right],
\ee
\be
{[}[[[\Pint,T],T],T],T]=4\pi\lambda\alpha^4\delta(0),
\ee
\be
{[}[[[[\Pint,T],T],T],T],T]=0.
\ee
From these commutators we construct the effective Hamiltonian, keeping
only terms which do not annihilate the vacuum and do not add more than
one zero mode,
\bea
\ob{\Pminus}&=& \sqrt{4\pi}\left[\mu^2\alpha+\frac16\lambda\alpha^3\right]
        \int\frac{dp^+}{\sqrt{p^+}}\Delta(p^+) a^\dagger(p^+) \\
  && +4\pi\left[\frac12\mu^2\alpha^2 +\frac{1}{24}\lambda\alpha^4\right]\delta(0). \nonumber
\eea
Figure~\ref{fig:phi4EffP} provides a graphical representation.
%
%

For a vacuum valence state, the valence eigenvalue problem is
\be
P_v\ob{\Pminus}|0\rangle={\cal E}^-\int dx^-|0\rangle,
\ee
with
\be
{\cal E}^-=\frac12\mu^2\alpha^2 +\frac{1}{24}\lambda\alpha^4.
\ee
The auxiliary equation (\ref{eq:aux}), projected onto the
one-zero-mode sector, yields
\be
\mu^2\alpha+\frac16\lambda\alpha^3=0.
\ee
These provide a complete match with the coherent-state result (\ref{eq:phi4var}),
with $\alpha=0$ or $\alpha^2=-6\mu^2/\lambda$, and $\alpha$
the vacuum expectation value for the field.

If we now consider the wrong-sign case, with $\mu^2\rightarrow-\mu^2$,
we find $\alpha=\pm\sqrt{6\lambda}/\mu$, which corresponds to the shift
of the field $\phi$ that brings the energy density to a minimum.
Thus, the inclusion of a zero mode in the LFCC $T$ operator allows
for the necessary shift in the field.  Also, as can be seen from
the commutator in (\ref{eq:PintT}), the effective Hamiltonian will
have terms that change the particle number by one and thereby
mix Fock states with odd and even numbers of particles,
which is characteristic of broken symmetry.

\section{Wick--Cutkosky model}
\label{sec:WC}

\begin{figure}[ht]
\vspace{0.2in}
\begin{center}
\begin{tabular}{c}
\includegraphics[width=12cm]{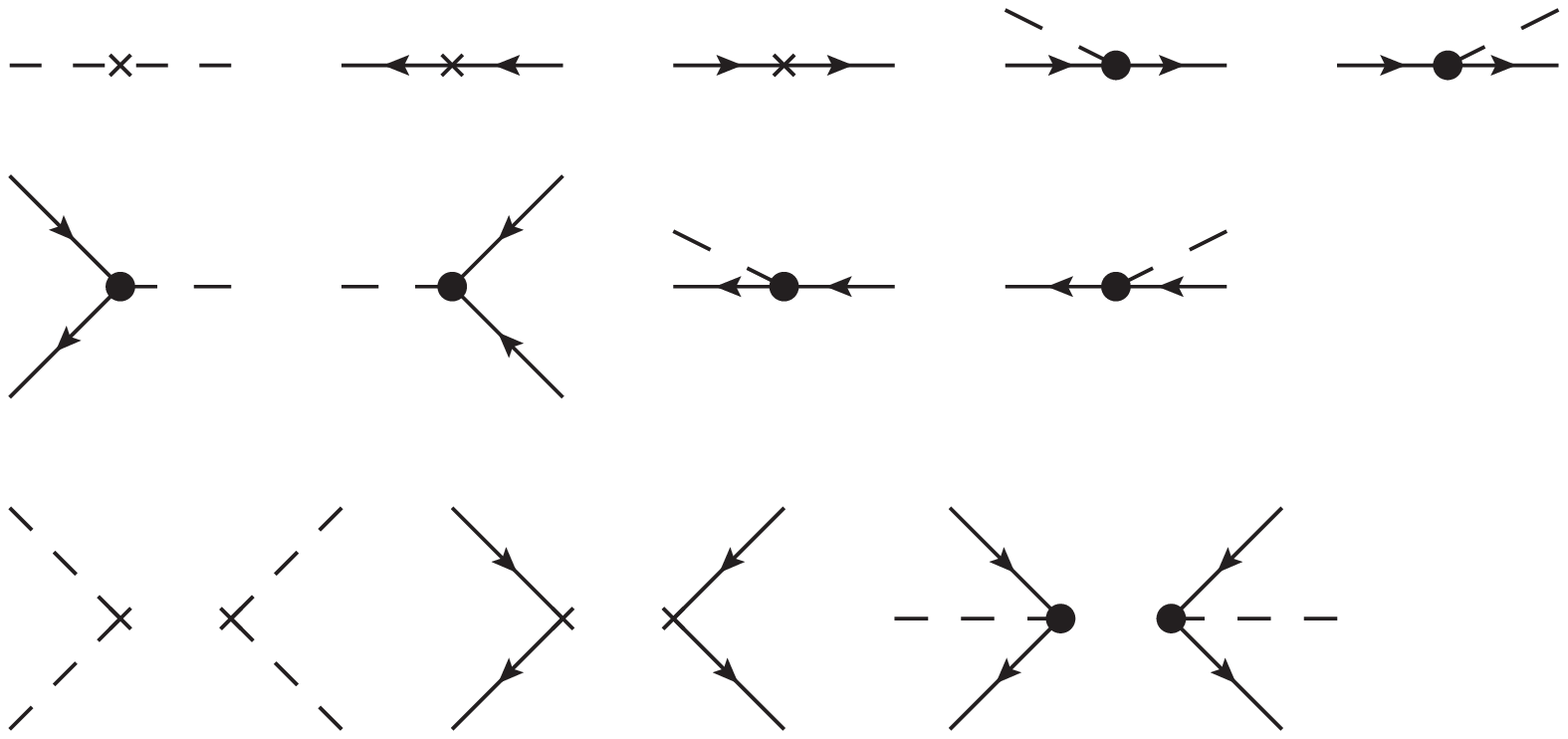} \\
(a) \\
\includegraphics[width=5cm]{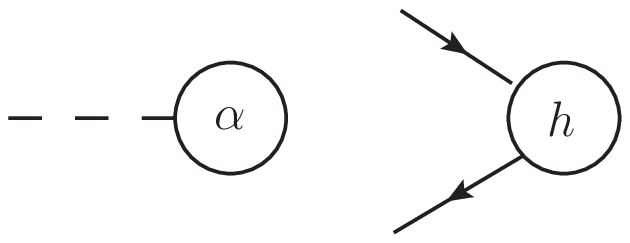} \\
(b)
\end{tabular}
\end{center}
\caption{\label{fig:WCPandT} 
Same as Fig.~\protect\ref{fig:phi3PandT}, but for the 
Wick--Cutkosky model.  The charged scalars are represented by
solid lines, with the arrow to the left (right) for
positive (negative) charge; the neutral scalars are represented
by dashed lines.  The zero-mode contributions are labeled by
$\alpha$ for the neutral scalar and by $h$ for the charged pair.
}
\end{figure}

To illustrate what happens in a more complicated theory,
we consider the Wick--Cutkosky model~\cite{WC} of a charged
scalar coupled to a neutral scalar.  The Lagrangian is
\be
{\cal L}=\frac12(\partial_\mu\phi)^2-\frac12\mu^2\phi^2
  +|\partial_\mu\chi|^2-m^2\chi^2-g\phi|\chi|^2,
\ee
where $\phi$ is the neutral scalar field and $\chi$ the
complex charged scalar field.  The Hamiltonian density is
\be
{\cal H}=\frac12 \mu^2 \phi^2+m^2|\chi|^2+g\phi|\chi|^2.
\ee
The mode expansion for the field $\phi$ is the same as
(\ref{eq:mode}) for $\phi^3$ theory; the mode expansion
for $\chi$ is
\be 
\chi=\int \frac{dp^+}{\sqrt{4\pi p^+}}
   \left\{ c_+(p^+)e^{-ip^+x^-/2} + c_-^\dagger(p^+)e^{ip^+x^-/2}\right\},
\ee
with $c^\dagger_\pm$ the creation operator for the positive
(negative) charge.  The nonzero commutation relation is
\be
[c_\pm(p^+),c_\pm^\dagger(\pp)]=\delta(p^+-\pp).
\ee
The free and interacting parts of the light-front Hamiltonian $\Pminus$ are
\be \label{eq:WCPfree}
\Pfree=\Pfreephi+\Pfreechi
\ee
with $\Pfreephi$ the free part for the $\phi$ field, as given in (\ref{eq:Pfree}),
\bea
\Pfreechi&=&\int dp^+ \frac{m^2}{p^+} 
      \left[c_+^\dagger(p^+)c_+(p^+) + c_-^\dagger(p^+)c_-(p^+)\right]  \\
 && +m^2\int \frac{dp_1^+ dp_2^+}{\sqrt{p_1^+ p_2^+}}
    \delta(p_1^++p_2^+)\left[c_+^\dagger(p_1^+)c_-^\dagger(p_2^+)+c_+(p_1^+)c_-(p_2^+)\right],
    \nonumber
\eea
and
\bea \label{eq:WCPint}
\Pint&=&g\int \frac{dp_1^+ dp_2^+ dp_3^+}{\sqrt{4\pi p_1^+ p_2^+ p_3^+}} 
         \delta(p_1^+ +p_2^+ +p_3^+) \\
  && \rule{0.5in}{0mm}\times \left[a^\dagger(p_1^+)c_+^\dagger(p_2^+)c_-^\dagger(p_3^+)
                 +a(p_1^+) c_+(p_2^+) c_-(p_3^+)\right] \nonumber \\
  && +g\int\frac{dp_1^+ dp_2^+}{\sqrt{4\pi p_1^+ p_2^+ (p_1^+ +p_2^+)}}
  \left\{a^\dagger(p_1^++p_2^+)c_+(p_1^+)c_-(p_2^+)\right. \nonumber \\
  &&  \rule{0.5in}{0mm} +a^\dagger(p_1^+)\left[c_+^\dagger(p_2^+)c_+(p_1^+ +p_2^+)
                              +c_-^\dagger(p_2^+)c_-(p_1^+ +p_2^+)\right] \nonumber \\
  && \rule{0.5in}{0mm} +\left[c_+^\dagger(p_1^+ +p_2^+)c_+(p_2^+)
                              +c_-^\dagger(p_1^+ +p_2^+)c_-(p_2^+)\right]a(p_1^+)
                         \nonumber \\
   && \rule{0.5in}{0mm} \left. +c_+^\dagger(p_1^+)c_-^\dagger(p_2^+)a(p_1^++p_2^+)\right\}. \nonumber
\eea
A graphical representation is given in Fig.~\ref{fig:WCPandT}.
%
%

We again focus on the zero-mode contributions and consider the $T$ operator
\be \label{eq:WCT}
T=T_\phi+T_\chi,
\ee
with $T_\phi$ defined as in (\ref{eq:T-Delta}) and
\be
T_\chi=\int dp_1^+ dp_2^+ 4 \pi\sqrt{p_1^+ p_2^+}h(p_1^+,p_2^+)
                         c_+^\dagger(p_1^+)c_-^\dagger(p_2^+).
\ee
The first term, $T_\phi$, creates a neutral-scalar zero mode; the
second creates a neutral pair of charged zero modes.  The function
$h$ is to be determined from the LFCC equations, but can be assumed to
be symmetric: $h(p_1^+,p_2^+)=h(p_2^+,p_1^+)$.
The first term in $T$ again corresponds to a shift of the field
$\phi$ by a constant $\alpha$.  

\begin{figure}[ht]
\vspace{0.2in}
\centerline{\includegraphics[width=12cm]{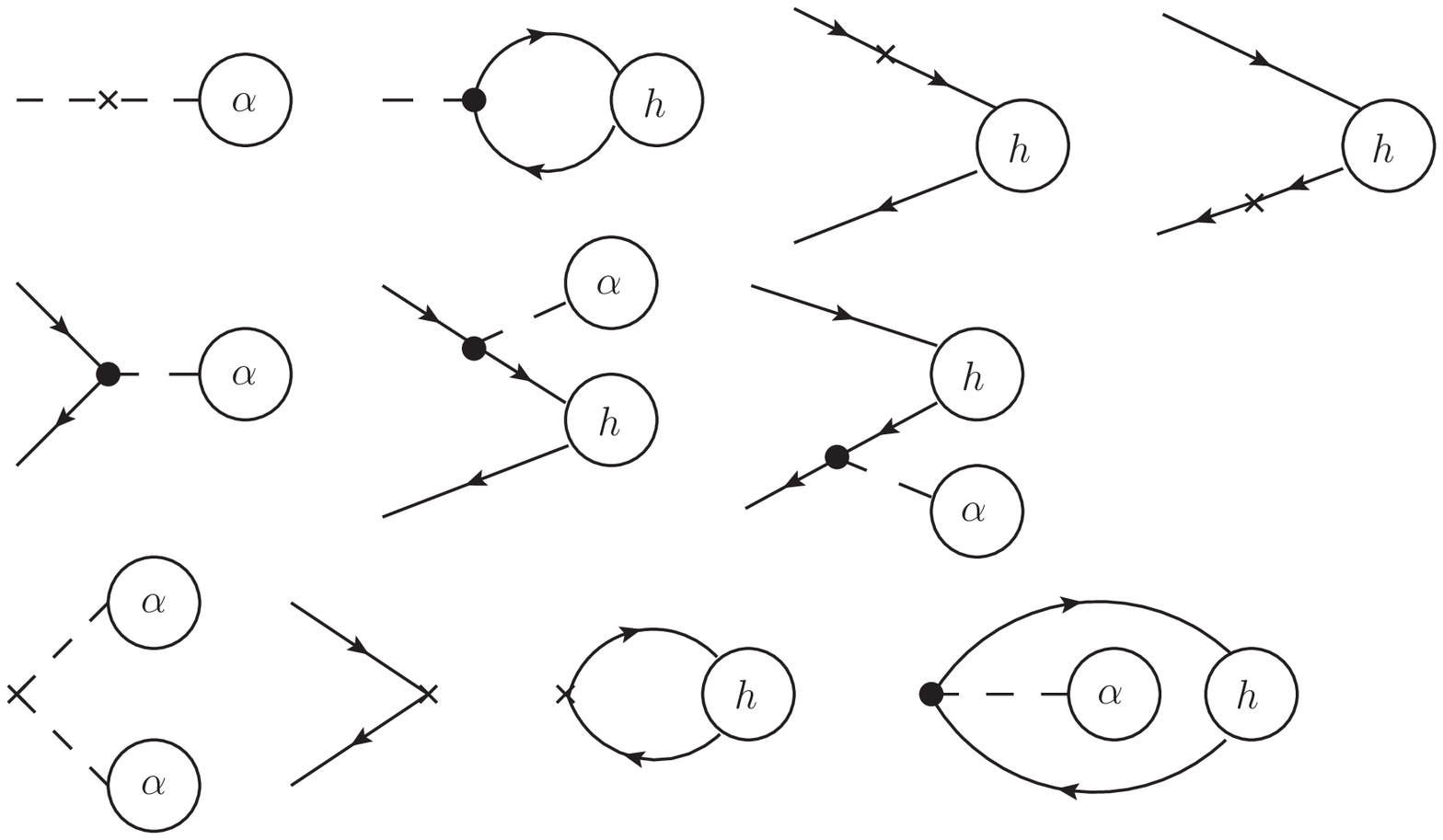}}
\caption{\label{fig:WCEffP} 
Same as Fig.~\protect\ref{fig:WCPandT}, but for the zero-mode
terms in the effective Hamiltonian $\ob{\Pminus}$.
}
\end{figure}

The relevant commutators with $\Pminus$ are given in
\ref{sec:WCcommutators}.
These yield an effective Hamiltonian of
\bea
\lefteqn{\ob{\Pminus}=\sqrt{4\pi}\alpha\mu^2\int dp^+ \frac{\Delta(p^+)}{\sqrt{p^+}}a^\dagger(p^+)
            +\frac12 \alpha^2\mu^2\int dx^-}&& \\
    && + \sqrt{4\pi}g\int\frac{dp_1^+ dp_2^+}{\sqrt{p_1^++p_2^+}}
                           h(p_1^+,p_2^+)a^\dagger(p_1^++p_2^+)   \nonumber \\
    && +(m^2+\alpha g)\int dp_1^+ dp_2^+
                 \left\{\rule{0mm}{0.25in}
                 4\pi\delta(p_1^++p_2^+)h(p_1^+,p_2^+)\right. \nonumber \\
    && \rule{0.5in}{0mm} +\frac{1}{\sqrt{p_1^+ p_2^+}}\left[\delta(p_1^++p_2^+)
         +4\pi(p_1^++p_2^+)h(p_1^+,p_2^+)\right]
            c_+^\dagger(p_1^+)c_-^\dagger(p_2^+) \nonumber \\
    && \rule{0.5in}{0mm} +(4\pi)^2\delta(p_1^++p_2^+)\int d\ppone d\pptwo \sqrt{\ppone \pptwo} \nonumber \\
    && \rule{2in}{0mm}  \left. \times
         h(p_1^+,\pptwo) h(p_2^+,\ppone) c_+^\dagger(\ppone)c_-^\dagger(\pptwo)
                   \rule{0mm}{0.25in}\right\}, \nonumber
\eea
where we list only those terms that do not annihilate the Fock vacuum
and do not create more than one neutral zero mode or one pair of
charged zero modes.
Figure~\ref{fig:WCEffP} shows a graphical representation.
%
%

In the vacuum valence sector, the eigenvalue problem
$P_v\ob{\Pminus}|0\rangle=P^-|0\rangle$ determines $P^-$ to be
\be \label{eq:WCP-}
P^-=\frac12\alpha^2\mu^2\int dx^-
   +4\pi(m^2+\alpha g)\int dp_1^+ dp_2^+ \delta(p_1^++p_2^+)h(p_1^+,p_2^+).
\ee
To have enough equations to solve for the unknowns $\alpha$ and $h(p_1^+,p_2^+)$
in the $T$ operator, we must project the auxiliary equation (\ref{eq:aux}) 
onto the Fock sector with one neutral zero mode and onto the sector
with a neutral pair.  These projections yield
\be \label{eq:firstprojection}
\alpha\mu^2\frac{\Delta(p^+)}{\sqrt{p^+}}+g\int_0^{p^+}\frac{d\pp}{\sqrt{p^+}}h(\pp,p^+-\pp)=0
\ee
and
\bea \label{eq:secondprojection}
\lefteqn{(m^2+\alpha g)\left[\frac{\delta(p_1^++p_2^+)}{\sqrt{p_1^+ p_2^+}}
   +4\pi\frac{p_1^++p_2^+}{\sqrt{p_1^+ p_2^+}}h(p_1^+,p_2^+)\right.}&& \\
  && \left. +(4\pi)^2\int d\ppone d\pptwo \delta(\ppone+\pptwo)\sqrt{p_1^+ p_2^+}
        h(\ppone,p_2^+) h(\pptwo,p_1^+)\rule{0mm}{0.3in} \right]=0. \nonumber
\eea
The last term in (\ref{eq:secondprojection}) is actually zero, as can be seen 
from the change of integration variables to $P^{\prime +}=\ppone+\pptwo$ and 
$x'=\ppone/P^{\prime +}$.  The integrals in this term then take the form
\be
\int P^{\prime +}dx' dP^{\prime +} \delta(P^{\prime +})
h(x'P^{\prime +},p_2^+)h((1-x')P^{\prime +},p_1^+),
\ee
which is proportional to $\int P^{\prime +}\delta(P^{\prime +})dP^{\prime +}=0$.

To solve these equations, we consider two cases, one where $m^2+\alpha g$ is not zero
and the other where it is.  For $m^2+\alpha g\neq0$, we have, from (\ref{eq:secondprojection})
\be
h(p_1^+,p_2^+)=-\frac{\delta(p_1^+ +p_2^+)}{4\pi(p_1^+ +p_2^+)}
   \longrightarrow -\frac{\Delta(p_1^+ +p_2^+)}{4\pi(p_1^+ +p_2^+)}.
\ee
Substitution into (\ref{eq:firstprojection}) yields
\be
\alpha \mu^2\frac{\Delta(p^+)}{\sqrt{p^+}}-g\frac{\Delta(p^+)}{\sqrt{p^+}}\int_0^{p^+}\frac{dp^{\prime +}}{4\pi p^+}=0
\ee
or
\be
\alpha=\frac{g}{4\pi\mu^2}.
\ee
We can also obtain an explicit form for the eigenvalue $P^-={\cal E}^-\int dx^-$
from (\ref{eq:WCP-}), with use of the same change of variables in the integration
\bea
\int dp_1^+ dp_2^+ \delta(p_1^++p_2^+)h(p_1^+,p_2^+)
&=&-\int P^+ dx dP^+ \delta(P^+) \frac{\delta(P^+)}{4\pi P^+}\\
  & =&\frac{-1}{4\pi}\delta(0)=\frac{-1}{(4\pi)^2}\int dx^-. \nonumber
\eea
This leaves
\be
{\cal E}^-=\frac12\alpha^2\mu^2-\frac{1}{4\pi}(m^2+\alpha g)
         =-\frac{1}{4\pi}\left(m^2+\frac{g^2}{8\pi\mu^2}\right).
\ee

When $m^2+\alpha g=0$, we obviously have
\be
\alpha=-\frac{m^2}{g}.
\ee
From (\ref{eq:WCP-}), we see that
\be
{\cal E}^-=\frac12\alpha^2\mu^2=\frac12\frac{m^4 \mu^2}{g^2},
\ee
independent of the form of $h(p_1^+,p_2^+)$.  The function $h$ does still
help determine the vacuum state and must still satisfy (\ref{eq:firstprojection}),
which becomes
\be
\int_0^{p^+}dp^{\prime +}h(p^{\prime +},p^+-p^{\prime +})=\frac{m^2\mu^2}{g^2}\Delta(p^+).
\ee
With a change of variable to $x'\equiv p^{\prime +}/p^+$, we obtain
\be
\int_0^1 dx' h(x'p^+,(1-x')p^+)=\frac{m^2\mu^2}{g^2}\frac{\Delta(p^+)}{p^+}.
\ee
Therefore, $h(x'p^+,(1-x')p^+)$ is proportional to $\Delta(p^+)/p^+$
and generally takes the form
\be
h(p_1^+,p_2^+)=\frac{m^2\mu^2}{g^2}f\left(\frac{p_1^+}{p_1^++p_2^+}\right) 
                  \frac{\Delta(p_1^++p_2^+)}{p_1^++p_2^+},
\ee
with $f$ any function that satisfies $\int_0^1f(x)dx=1$.  In the limit
that $\Delta(p^+)\rightarrow\delta(p^+)$, the form of $f$ is irrelevant.

From the form of $P^-$ in (\ref{eq:WCP-}), we see that, as expected for
a cubic theory~\cite{Baym}, the spectrum is unbounded from below.
However, as in the case of $\phi^3$ theory, the LFCC auxiliary equations
determine only local extrema.

\section{Summary}
\label{sec:summary}

We have considered various examples of two-dimensional scalar
theories where zero modes can play a role in the calculation
of light-front Hamiltonian eigenstates.  In each case, the LFCC method
is able to incorporate the zero modes in a sensible way and, where it
can be compared with a coherent-state analysis, obtains equivalent
results for the local extrema of the energy density and the
vacuum expectation value for the bosonic field.
To do this, the $T$ operator must include terms that allow creation
of modes with infinitesimal longitudinal momentum,
with the limit of zero momentum taken at a later stage
in the calculation.  The vacuum state is then a generalized
coherent state of zero modes, created from the trivial Fock
vacuum by the operator $e^T$.

Although the examples are limited to two dimensions, there
is no particular restriction on a direct extension to three
or four dimensions.  The zero-mode terms would include a
dependence on transverse momenta.  

With these tools in place, one can use the LFCC method to
explore symmetry breaking nonperturbatively.  A calculation in
$\phi^4$ theory that includes as many as four zero modes in the
$T$ operator should be sufficient to compute the critical
coupling for dynamical symmetry breaking; this would parallel
the calculations done in equal-time quantization~\cite{CC-QFT}.
Another accessible application is a nonperturbative calculation
of the Higgs mechanism and the associated breaking of a 
continuous symmetry.  The general aim is, of course, to
apply these methods to the nonperturbative solution of
QCD in terms of hadronic wave functions, particularly
with respect to symmetry-breaking effects.

\acknowledgments
This work was supported in part by the U.S. Department of Energy
through Contract No.\ DE-FG02-98ER41087.

\appendix

\section{Commutators for the Wick--Cutkosky model} \label{sec:WCcommutators}

The commutators needed to construct the effective Hamiltonian for the
Wick--Cutkosky model are below, with $\Pfree$ defined in (\ref{eq:WCPfree}),
$\Pint$ in (\ref{eq:WCPint}), and $T$ in (\ref{eq:WCT}).
The commutators of $\Pfreephi$ with $T_\phi$ are
the same as those for $\phi^3$ theory, with $g(p^+)=\alpha\Delta(p^+)$,
given in Eqs.~(\ref{eq:PfreeT}) and (\ref{eq:PfreeT2}).
The other commutators are
\bea
[\Pfreechi,T_\chi]&=&4\pi m^2\int dp_1^+ dp_2^+\left\{\rule{0mm}{0.3in}
           \delta(p_1^+ +p_2^+)h(p_1^+,p_2^+)         \right. \\
  && + \frac{p_1^+ +p_2^+}{\sqrt{p_1^+ p_2^+}}h(p_1^+,p_2^+)
                               c_+^\dagger(p_1^+)c_-^\dagger(p_2^+) \nonumber \\
  && +\delta(p_1^+ +p_2^+)\int d\pp 
      \left[ \sqrt{\frac{\pp}{p_1^+}}h(\pp,p_2^+)c_+^\dagger(\pp)c_+(p_1^+) 
                                     \right.  \nonumber \\
  && \left.\left. \rule{1.2in}{0mm}
        +\sqrt{\frac{\pp}{p_2^+}}h(p_1^+,\pp)c_-^\dagger(\pp)c_-(p_2^+)
                         \right]\right\},           \nonumber
\eea
\be
[[\Pfreechi,T_\chi],T_\phi]=0,
\ee
\be
[[\Pfreechi,T_\phi],T_\chi]=0,
\ee
\bea
[[\Pfreechi,T_\chi],T_\chi]&=&2(4\pi)^2m^2\int dp_1^+ dp_2^+\delta(p_1^++p_2^+)
     \int d\ppone d\pptwo \\
  && \times \sqrt{\ppone \pptwo}h(p_1^+,\pptwo)h(p_2^+,\ppone)
     c_+^\dagger(\ppone)c_-^\dagger(\pptwo), \nonumber
\eea
\be
[[[\Pfreechi,T_\chi],T_\chi],T_\chi]=0,
\ee
\be
[\Pint,T_\phi]=\frac{\alpha g}{m^2}\Pfreechi,
\ee
\be
[[\Pint,T_\phi],T]=\frac{\alpha g}{m^2}[\Pfreechi,T_\chi],
\ee
\be
[[[\Pint,T_\phi],T],T]=\frac{\alpha g}{m^2}[[\Pfreechi,T_\chi],T_\chi],
\ee
\be
[[[[\Pint,T_\phi],T],T],T]=0,
\ee
\bea
\lefteqn{[\Pint,T_\chi]=g\sqrt{4\pi}\int dp_1^+ dp_2^+ dp_3^+\delta(p_1^+ +p_2^+ +p_3^+)
   \left[\rule{0mm}{0.3in} h(p_2^+,p_3^+) \right.} && \\
   && \rule{1in}{0mm}
   +\int d\ppone\sqrt{\frac{\ppone}{p_2^+}}h(\ppone,p_3^+)c_+^\dagger(\ppone)c_+(p_2^+) 
                  \nonumber \\
   && \left.\rule{1in}{0mm}
     +\int d\pptwo\sqrt{\frac{\pptwo}{p_3^+}} h(p_2^+,\pptwo) c_-^\dagger(\pptwo)c_-(p_3^+)
        \right]   \frac{a(p_1^+)}{\sqrt{p_1^+}} \nonumber \\
   && +g\sqrt{4\pi}\int dp_1^+ dp_2^+ \left\{
              \frac{h(p_1^+,p_2^+)}{\sqrt{p_1^++p_2^+}}a^\dagger(p_1^++p_2^+) 
                                       \right. \nonumber \\
   &&  \rule{0.5in}{0mm} + \int d\pptwo \sqrt{\frac{\pptwo}{p_1^+p_2^+}}
             h(p_1^++p_2^+,\pptwo)a^\dagger(p_1^+)c_+^\dagger(p_2^+)c_-^\dagger(\pptwo)
                 \nonumber \\
   && \rule{0.5in}{0mm} +\int d\ppone \sqrt{\frac{\ppone}{p_1^+p_2^+}}
             h(p_1^++p_2^+,\ppone)a^\dagger(p_1^+)c_+^\dagger(\ppone)c_-^\dagger(p_2^+)
                 \nonumber \\
   && \rule{0.5in}{0mm} +\int d\pptwo \sqrt{\frac{\pptwo}{p_1^+(p_1^++p_2^+)}}
             h(p_2^+,\pptwo)c_+^\dagger(p_1^++p_2^+)c_-^\dagger(\pptwo)a(p_1^+)
                 \nonumber \\
   && \rule{0.5in}{0mm} +\int d\ppone \sqrt{\frac{\ppone}{p_1^+(p_1^++p_2^+)}}
             h(p_2^+,\ppone)c_+^\dagger(\ppone)c_-^\dagger(p_1^++p_2^+)a(p_1^+)
                 \nonumber \\
   && \rule{0.5in}{0mm} +\int d\ppone \sqrt{\frac{\ppone}{p_1^+(p_1^++p_2^+)}}
             h(p_2^+,\ppone)a^\dagger(p_1^++p_2^+)c_+^\dagger(\ppone)c_+(p_1^+)
                 \nonumber \\
   && \rule{0.5in}{0mm} \left. +\int d\pptwo \sqrt{\frac{\pptwo}{p_2^+(p_1^++p_2^+)}}
             h(p_1^+,\pptwo)a^\dagger(p_1^++p_2^+)c_-^\dagger(\pptwo)c_-(p_2^+)
                      \right\},  \nonumber
\eea
\bea
[[\Pint,T_\chi],T_\phi]&=&4 \pi g\alpha \int dp_2^+ dp_3^+\delta(p_2^+ +p_3^+)
  \left\{ \rule{0mm}{0.3in}h(p_2^+,p_3^+) \right. \\
  &&  \rule{0.25in}{0mm}
   +\int dp_1^+\sqrt{\frac{p_1^+}{p_2^+}}h(p_1^+,p_3^+)
     \left[c_+^\dagger(p_1^+)c_+(p_2^+) \right. \nonumber \\
   &&  \rule{2in}{0mm} \left. \left. + c_-^\dagger(p_1^+)c_-(p_2^+)\right]
        \rule{0mm}{0.3in} \right\}  \nonumber \\
   && +4 \pi g\alpha \int dp_1^+ dp_2^+ \frac{p_1^++p_2^+}{\sqrt{p_1^+ p_2^+}}
           h(p_1^+,p_2^+) c_+^\dagger(p_1^+) c_-^\dagger(p_2^+), \nonumber
\eea
\bea
[[\Pint,T_\chi],T_\chi]&=&2g(4\pi)^{3/2}\int dp_1^+ dp_2^+ dp_3^+
    \delta(p_1^++p_2^+ +p_3^+) \\
    && \int d\ppone d\pptwo \sqrt{\frac{\ppone\pptwo}{p_1^+}} \nonumber \\
    && \rule{0.5in}{0mm} \times  h(\ppone,p_3^+)h(\pptwo,p_2^+)
                   c_+^\dagger(\ppone) c_-^\dagger(\pptwo)a(p_1^+) \nonumber \\
    && +2 g (4\pi)^{3/2}\int \frac{dp_1^+ dp_2^+}{\sqrt{p_1^++p_2^+}}
      \int d\ppone d\pptwo \sqrt{\ppone\pptwo}\nonumber \\
    && \rule{0.5in}{0mm} \times  h(p_1^+,\pptwo) h(p_2^+,\ppone) 
         a^\dagger(p_1^++p_2^+) c_+^\dagger(\ppone) c_-^\dagger(\pptwo), \nonumber
\eea
\bea
[[[\Pint,T_\chi],T_\phi],T_\chi]&=&[[[\Pint,T_\chi],T_\chi],T_\phi] \\
&=&2g\alpha(4\pi)^2
   \int dp_1^+ dp_2^+ dp_3^+  \delta(p_2^+ +p_3^+) \\
   && \int d\pptwo \sqrt{p_1^+\pptwo}
     h(p_1^+,p_3^+) h(p_2^+,\pptwo) c_+^\dagger(p_1^+) c_-^\dagger(\pptwo), \nonumber
\eea
and 
\be
[[[[\Pint,T_\chi],T_\phi],T_\chi],T]=0.
\ee
These are then combined according to the Baker--Hausdorff expansion to
construct the effective Hamiltonian.


\end{document}